# Ultra strong and ductile eutectic high entropy alloy fabricated by selective laser melting

*Fan Yang, Lilin Wang\*, Zhijun Wang\*, Qingfeng Wu, Kexuan Zhou,*

*Weidong Huang, Xin Lin*

*State Key Laboratory of Solidification Processing, Northwestern Polytechnical*

*University, Xi'an 710072, China*

**Abstract:** With important application prospects, eutectic high entropy alloys have received extensive attention for their excellent strength and ductility in a large temperature range. The excellent casting characteristics of eutectic high entropy alloys make it possible to achieve well manufacturability of selective laser melting. For the first time, we have achieved crack-free eutectic high entropy alloy fabricated by selective laser melting, which has excellent mechanical properties in a wide temperature range of -196°C ~ 760°C due to ultra-fine eutectic lamellar spacing of 150 ~ 200nm and lamellar colony of 2 ~ 6μm. Specifically, the room temperature tensile strength exceeds 1400MPa and the elongation is more than 20%, significantly improved compared with those manufactured by other techniques with lower cooling rate.

Additive manufacturing can precisely fabricate complex geometrical structures and high-performance components. The selective laser melting (SLM) is a representative additive manufacturing technique based on powder bed fusion with particular advantages [1-4]. In SLM, the solidification process in the micro-molten pool occurring at extremely rapid cooling rate can greatly refine the microstructure to possess high performance of mechanical properties. However, the residual stress during SLM also causes serious cracking problems, which hinders its engineering application [5-8]. In order to solve the problem, conventional methods by optimizing the process parameters or optimizing the strength and ductility matching have been used to reduce the crack sensitivity of the alloy [4, 6, 7, 9-11]. The emergence of new kinds of alloys, such as high entropy alloys (HEAs), provides new alloy candidates for SLM.

\* Corresponding author. wlilin@nwpu.edu.cn
\* Corresponding author. zhjwang@nwpu.edu.cn

HEAs shape a new class of alloys in the way of multiple principal elements [12]. A large number of HEAs with excellent ductility have achieved crack-free processing by additive manufacturing, but their tensile strength needs to be further improved [13-20]. Eutectic high entropy alloys (EHEAs) have an in-situ composite structure with good combination of strength and ductility. As eutectic alloys, EHEAs also have excellent castability [21-26]. The solidification behaviors of eutectic alloys are believed to be suitable for additive manufacturing. Especially, the narrow solidification temperature range is beneficial to reduce the hot crack in additive manufacturing. In recent years, EHEAs have been achieved good forming quality in laser direct energy deposition (DED), but with a lower tensile yield strength (<800MPa) at room temperature [27, 28]. Compared with DED, SLM has much faster cooling rate in the process of ultrafast solidification. It is expected to obtain finer eutectic structure and improve mechanical properties. It is also intriguing to explore how the fine microstructure of EHEAs evolves. However, there are few reports on SLM of eutectic alloys. Based on a EHEA ($Ni_{30}Co_{30}Cr_{10}Fe_{10}Al_{18}W_1Mo_1$ (at. %)) with excellent strength and ductility, we achieve good forming quality with the high relative density SLM samples. The yield strength is around 1000 MPa and tensile strength is more than 1400 MPa with more than 20% uniform tensile elongation at room temperature.

The pre-alloyed EHEA spherical powder (15 ~ 53μm in size) was prepared by gas atomization for SLM. The powder was dried in a vacuum furnace at 120°C for 4 h to remove moisture before SLM. SLM experiment was performed by using a BLT S210 machine. The processing parameters of SLM are shown in Table.S1, and interlayer scanning angle is 67°. The chemical composition of the powder and the as-built sample were measured by ICP, and the results are shown in Table S2. The as-built samples were annealed at 900°C for 2h followed by air cooling to relax the residual stress. The surface defects of the polished sample were observed by an optical microscopy (OM, Keyence VHX-2000). Porosity and interlamellar spacing were measured using image processing software (Image-Pro Plus). Secondary electron (SE) imaging, and electron backscatter diffraction (EBSD) were carried out on a TESCAN MIRA3. XRD analyses were determined using an X-ray diffractometer (Shimadzu, MAXima XRD-7000) with Cu Kα radiation at a scanning rate of $6° \text{ min}^{-1}$, and a 2θ angle ranging from 30° to 100°. The crack-free samples with dimensions of $52 \times 13 \times 5 \text{ mm}^3$ (5mm is the height) were manufactured with an optimized volume energy density (VED) of 50~80 $J/mm^3$. Tensile tests at room temperature, low temperature and elevated temperature were carried out on annealed EHEA samples with dog-bone-shaped samples. The tensile tests with a strain rate of $1 \times 10^{-3} \text{ s}^{-1}$ were performed at -196°C, 650°C, 760°C and 980°C, respectively. The tensile direction of the sample is perpendicular to the building direction.

Firstly, the processing parameters were evaluated by characterizing defects with respect to the VED in a wide range, as shown in Fig. 1. Fig. 1a is the statistical results of porosity of EHEA samples fabricated at different process parameters. With low VED, irregular pores appear in lack of fusion due to the low molten pool temperature and poor melt fluidity, as shown in Fig. 1b. With the increase of VED, there is a stable range for low porosity less than 0.05% within the VED range of 50 ~ 80 $J/mm^3$, and the typical polished morphology is shown in Fig. 1c. When the VED is too high, regular circular pores with keyhole defects appear at the bottom of deep and narrow molten pool, as shown in Fig. 1d [29].

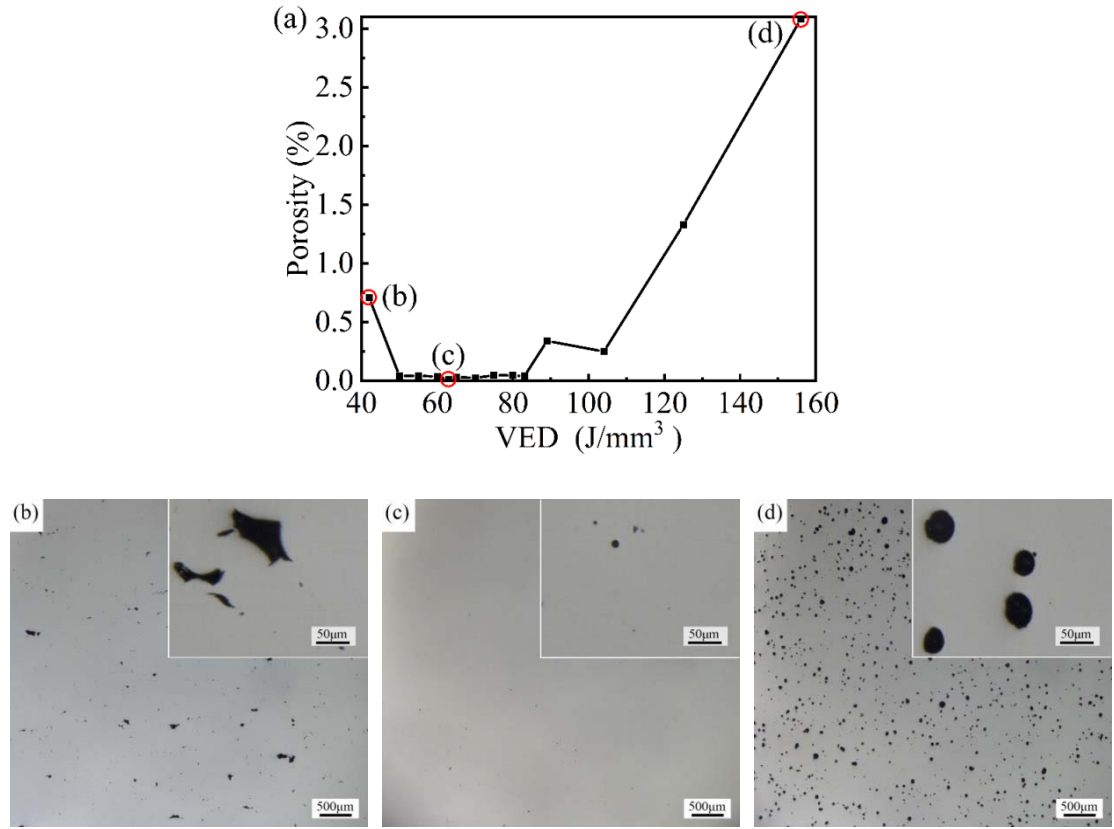

Fig. 1. (a) Porosity of EHEA samples fabricated at different process parameters; (b, c, d) optical graphs of EHEA samples to show the defect distributions with VED of 42 J/mm$^3$, 63 J/mm$^3$, 156 J/mm$^3$, respectively.

The XRD analysis of EHEA samples fabricated by SLM is shown in Fig. S2 to confirm the phase selection of FCC and B2 dual phases [30]. For low porosity samples, we further characterized the microstructure by focusing on the fine lamellar in rapid solidification. Fig. 2a and 2b are SEM images of cross plane and longitudinal plane of as-built EHEA samples with VED of 83 J/mm$^3$ (laser power of 200W). It can be seen in Fig. 2a$_1$ that the cross plane of the sample has an interlaced molten pool boundary line, indicated by the yellow line reflecting the scanning strategy of interlayer scanning angle of 67°. Enlarged image in Fig. 2a$_2$ shows that the cross plane is full of equiaxed colony. In Fig. 2b$_1$, a clear morphology of the molten pool was observed on the longitudinal plane, as shown by the yellow line. The molten pool showed directional growth characteristics of microstructures as shown in Fig. 2b$_2$, but did not go through the cladding layer. The Fig. 2a$_3$ and 2b$_3$ with high magnification clearly show total eutectic structure, most of which are regular lamellar eutectic structure, and a small amount of irregular eutectic structure exists near the molten pool boundary. Moreover, the fine interlamellar spacing was measured as 150 ~ 200nm, about 5 times smaller than that of other EHEA alloys fabricated by DED [28]. This is because the cooling rate of SLM is as fast as $1 \times 10^7$ °C/s, which is two orders of magnitude higher than DED [2]. The eutectic layer spacing is inversely proportional to the solidification rate, so the eutectic lamellar structure of EHEA obtained by SLM is extremely fine. The as-built EHEA sample microstructures at other process parameters are shown in Fig. S3. It reveals that the SLM process parameters at different VED and laser powers have almost the same eutectic structure.

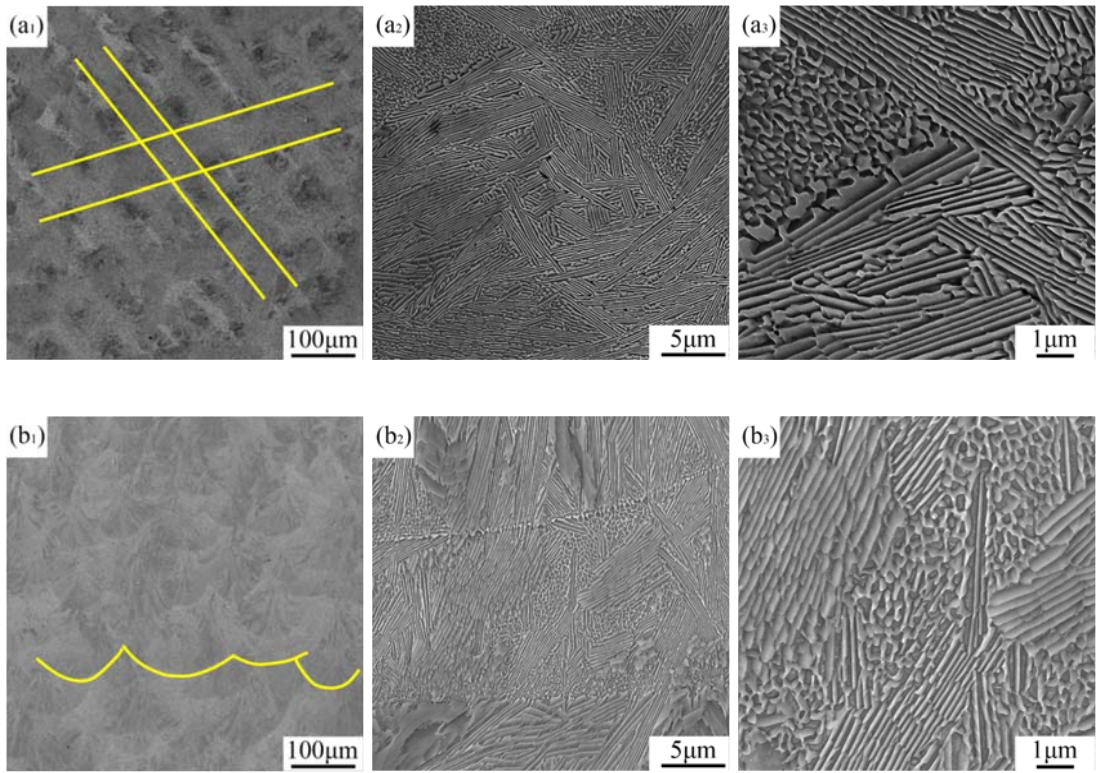

Fig. 2.SEM microstructures of as-built EHEA with VED of 83 J/mm$^3$ (laser power of 200W) from cross plane ($a_1$, $a_2$, $a_3$,), and longitudinal plane ($b_1$, $b_2$, $b_3$,).

In Fig.2, it shows that the lamellar colony is also very small, even along the building direction. We further analyzed the colony distribution with EBSD along the building direction, as shown in Fig.3. Fig. 3a presents the inverse pole figure (IPF) map of FCC phase of as-built EHEA sample with VED of 83 J/mm$^3$ (laser power of 200W). The colonies are composed of equiaxed grains and columnar grains. There are a large number of equiaxed grains at the bottom of the molten pool while a large number of columnar grains in the molten pool, growing along the normal direction of the molten pool boundary. The distribution of lamellar colony size is shown in Fig. 3b, and most lamellar colonies are in the size range of 2 ~ 6μm.

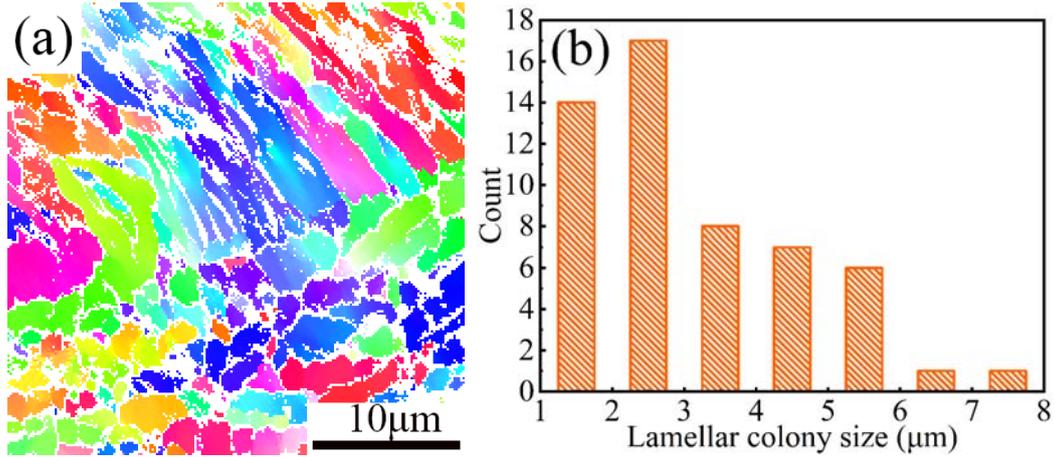

Fig. 3. (a) IPF map of FCC phase of as-built EHEA with VED of 83 J/mm$^3$ (laser power of 200W); (b) distribution of lamellar colony size.

The ultra-fine lamellar eutectic structure and smaller colony are beneficial to the strength and ductility. We further investigated the mechanical behavior in a wide temperature range from -196°C to 980°C. The EHEA tensile samples were carried out stress relief annealing in the consideration of the large stress in as-built samples as shown in Fig. S4. Fig. S5 presented SEM microstructures of the annealed EHEA sample with VED of 65 J/mm$^3$ (laser power of 150W). There was no obvious change in the microstructure before and after heat treatment compared with Fig.2.

The tensile results are shown in Fig. 4. Fig. 4a shows the room-temperature tensile stress-strain curve of EHEA samples with different SLM process parameters. It can be seen that in a large range of SLM process parameter window with good formability, the mechanical properties of EHEA samples fabricated by different process parameters are similar and the dispersion of results on mechanical properties is quite small. At room temperature, the average yield strength ($\sigma_{0.2}$) of the EHEA sample reached 995 MPa, the average tensile strength reached 1402 MPa, with an average uniform elongation of 18%. By comparing with the yield strength of HEAs fabricated by others additive manufacturing (including DED [16, 17, 28], SLM [13-15], wire and arc additive manufacturing (WAAM) [18]) in references at room temperature tensile, the EHEA samples fabricated by SLM have excellent combined mechanical properties of strength and ductility. The ultra strong and ductility are attributed to the contribution of ultra-fine lamellar eutectic structure to strength and the contribution of smaller eutectic colonies to ductility, as shown in Fig. 4b.

Tensile tests were also carried out at cryogenic temperature and elevated temperatures. The results are shown in Fig. 4c and detailed tensile test results of the EHEA samples are shown in Table S2. At cryogenic tensile tests of -196°C, the average yield strength ($\sigma_{0.2}$) is 1221 MPa, and the average tensile strength is up to 1545 MPa. At tensile tests at 650°C, the average yield strength ($\sigma_{0.2}$) of the EHEA sample reaches 592 MPa, and the average tensile strength is 714 MPa. After further increasing the tensile tests temperature to 760°C, the average yield strength ($\sigma_{0.2}$) and the tensile strength decrease seriously to be 214 MPa and 287 MPa respectively. Superplastic deformation occurred, and the elongation after fracture exceeded 300 % when tensile tests were performed at 980°C, but the strength was as low as 31 MPa. The variation of yield strength with temperature is represented in Fig. 4d. The yield strength decreases rapidly after 650°C, due to the ductile-brittle transition temperature of B2 phase around 750°C [31-33].

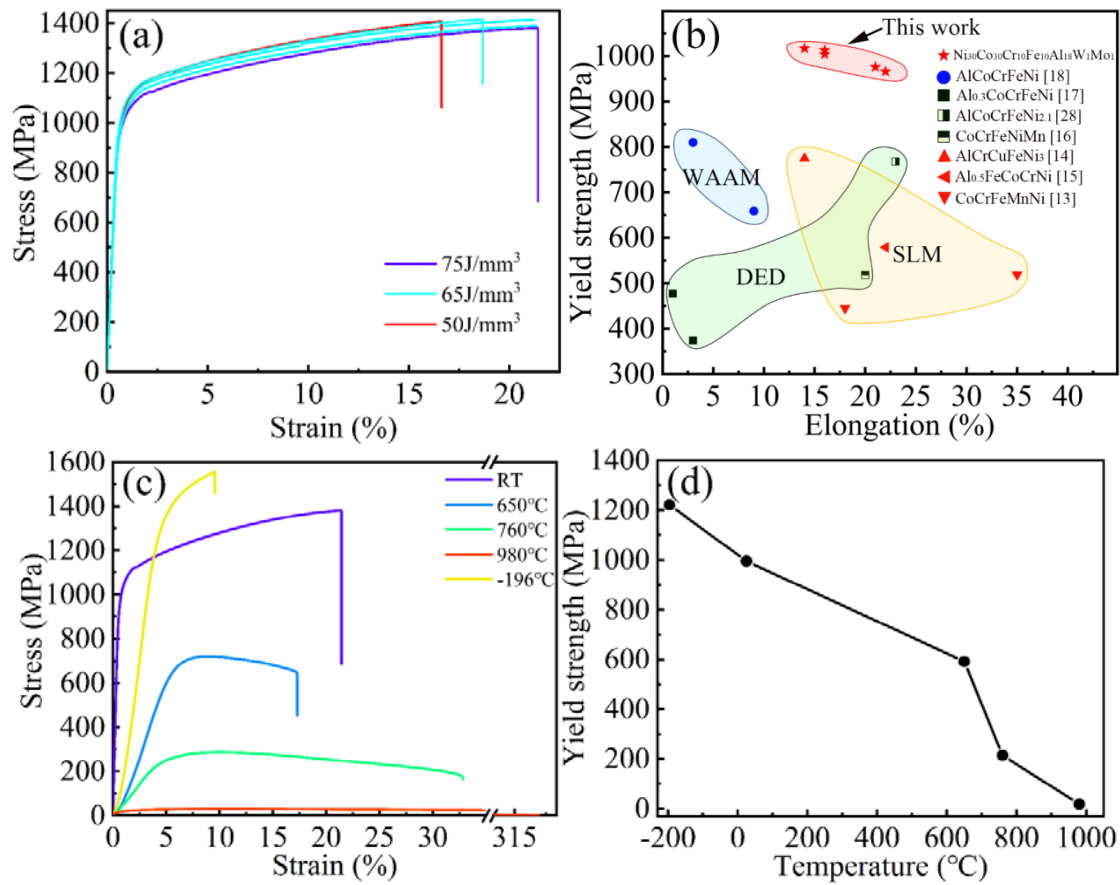

Fig. 4. (a) Tensile stress-strain curve at room temperature of EHEA samples with different VED; (b) comparison of yield strength with different additive manufacturing processes of HEAs [13-18, 28]; (c) tensile stress-strain curve of EHEA samples at different temperature; (d) the variation of yield strength with temperature.

In summary, we explored a wide range of SLM process parameters for a EHEA $Ni_{30}Co_{30}Cr_{10}Fe_{10}Al_{18}W_1Mo_1$ with excellent mechanical properties. In the energy density range of 50 ~ 80 J/mm³, the porosities of the as-built samples are less than 0.05% without any micro-crack. Interestingly, the EHEA samples fabricated by SLM have total eutectic structure with ultra-fine lamellar spacing of 150 ~ 200nm within the small colony with size of 2 ~ 6μm. The refinement of lamellar eutectic structure results in excellent mechanical properties. The yield and ultimate tensile strength at room temperature is around 1.0GPa, and 1.4GPa respectively with uniform elongation larger than 18%. At the same time, it has excellent low- and medium- temperature tensile mechanical properties from -196°C to 760°C.

**Acknowledgements**

Fan Yang gives his thanks to Zishu Chai for her help on the experiments. This work was supported by the "National Key R&D Program of China (2018YFC0310400)" and the Research Fund of the State Key Laboratory of Solidification Processing, China (Grant No. 2020-TS-06, 2021-TS-02).

# Supplementary materials

Table S1. The process parameters of SLM for EHEA ($Ni_{30}Co_{30}Cr_{10}Fe_{10}Al_{18}W_1Mo_1$ (at. %)).

| Laser power (W) | Scanning velocity (mm/s) | Hatch spacing (μm) | Layer thickness (μm) | Volume energy density (J/mm$^3$) |
|---|---|---|---|---|
| 150 | 400 | | | 156 |
| | 500 | | | 125 |
| | 600 | | | 104 |
| | 700 | | | 89 |
| | 750 | | | 83 |
| | 781 | | | 80 |
| | 800 | | | 78 |
| | 833 | | | 75 |
| | 893 | | | 70 |
| | 962 | 80 | 30 | 65 |
| | 1000 | | | 63 |
| | 1042 | | | 60 |
| | 1136 | | | 55 |
| | 1250 | | | 50 |
| | 1500 | | | 42 |
| 200 | 800 | | | 104 |
| | 1000 | | | 83 |
| | 1600 | | | 69 |
| | 2000 | | | 42 |

The volume energy density (VED) is defined as follows:

$$VED = \frac{P}{vht}$$

where $P$ is the laser power (W), $v$ the scan speed (mm/s), $h$ the hatching space (μm), and $t$ the layer thickness (μm).

Table S2. Chemical compositions of EHEA powder and as-built EHEA samples (wt. %).

| Element | Ni | Co | Cr | Fe | Al | W | Mo |
|---|---|---|---|---|---|---|---|
| Powder | Bal. | 32.70 | 10.01 | 10.78 | 9.30 | 1.55 | 0.52 |
| As-built | Bal. | 33.88 | 10.00 | 11.50 | 8.96 | 1.52 | 0.81 |

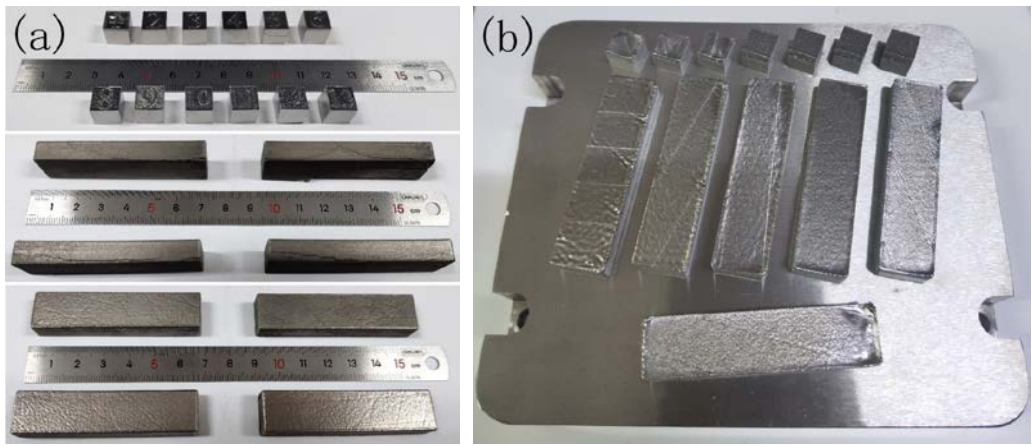

Fig. S1. The pictures of as-built EHEA samples.

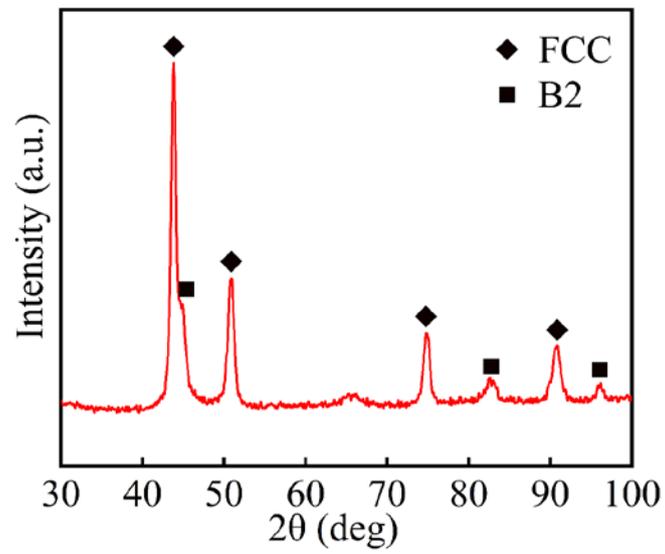

Fig. S2. XRD patterns of as-built EHEA sample.

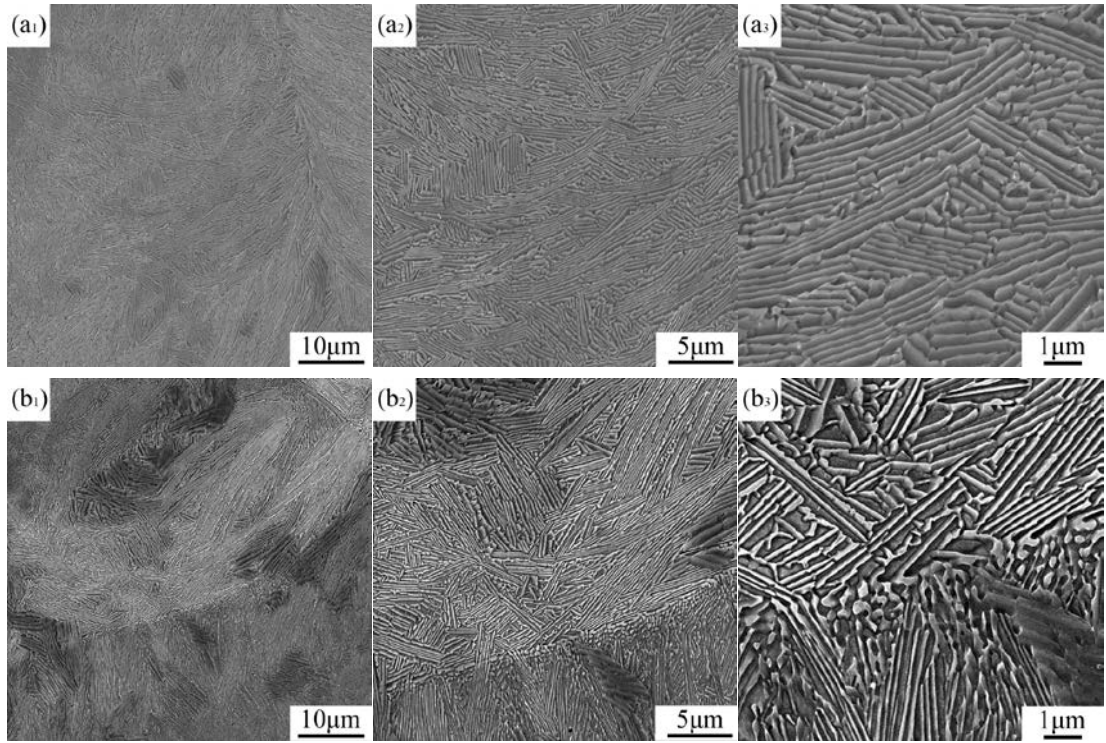

Fig. S3 SEM micrographs of as-built EHEA; ($a_1$, $a_2$, $a_3$) from longitudinal plane and VED is 156 J/mm$^3$ (laser power of 150W); ($b_1$, $b_2$, $b_3$) from cross plane and VED is 83 J/mm$^3$.

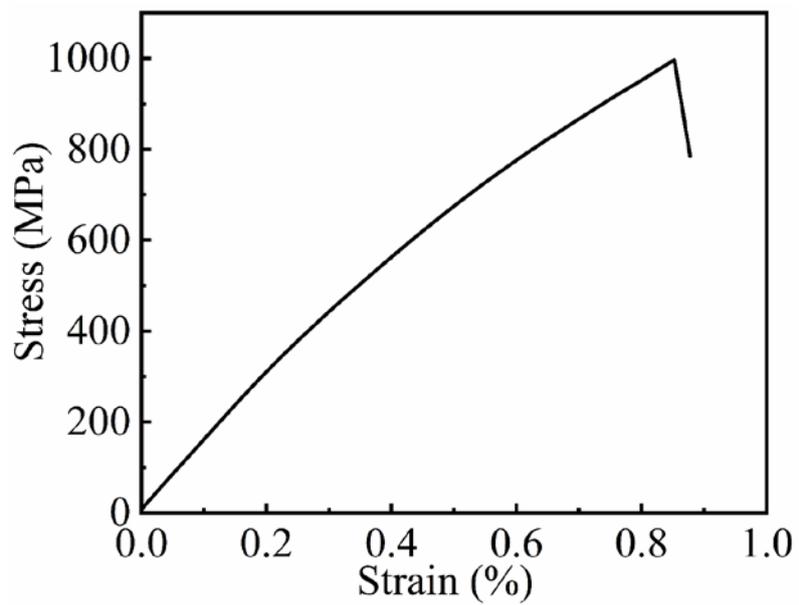

Fig. S4 Tensile stress-strain curve at room temperature of as-built EHEA samples with VED of 83 J/mm$^3$.

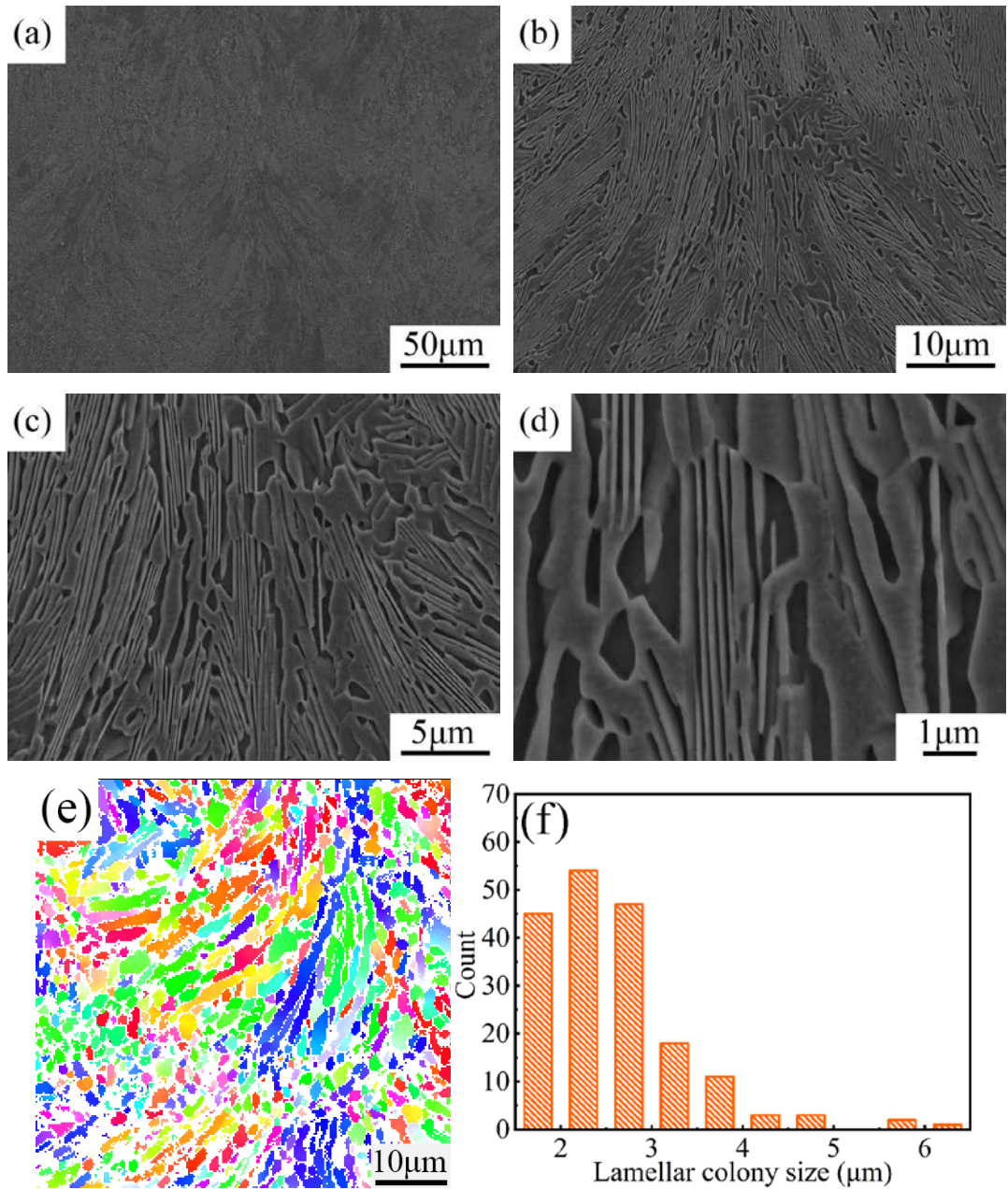

Fig. S5 (a, b, c, d) SEM micrographs of the annealed EHEA sample with VED of 65 J/mm$^3$ (laser power of 150W); (e) IPF map of FCC phase of the annealed EHEA sample with VED of 65 J/mm$^3$ (laser power of 150W); (f) distribution of lamellar colony size.

Table S3. Mechanical properties of Tensile tests

| Temperature | VED (J/mm$^3$) | Yield strength ($\sigma_{0.2}$, MPa) | Tensile strength (MPa) | Elongation (%) |
|---|---|---|---|---|
| RT | 50 | 1013 | 1408 | 16 |
| RT | 65 | 1004 | 1414 | 16 |
| RT | 65 | 976 | 1390 | 21 |
| RT | 65 | 1017 | 1416 | 14 |
| RT | 75 | 966 | 1381 | 22 |
| 650°C | 50 | 585 | 721 | 14 |
| 650°C | 65 | 598 | 706 | 15 |
| 760°C | 75 | 214 | 287 | 29 |
| 980°C | 83 | 18 | 31 | >300 |
| -196°C | 60 | 1191 | 1550 | 6 |
| -196°C | 60 | 1250 | 1540 | 6 |